# Commensurate-incommensurate transition for graphene on hexagonal boron nitride

C. R. Woods[1], L. Britnell[1], A. Eckmann[2], R. S. Ma[3], J. C. Lu[3], H. M. Guo[3], X. Lin[3], G. L. Yu[1], Y. Cao,[4] R. V. Gorbachev[4], A.V. Kretinin[1], J. Park[1,5], L. A. Ponomarenko[1], M. I. Katsnelson[6], Yu. N. Gornostyrev[7], K. Watanabe[8], T. Taniguchi[8], C. Casiraghi[2], H. -J. Gao[3], A. K. Geim[4], K. S. Novoselov[1]

[1]School of Physics and Astronomy, University of Manchester, Manchester M13 9PL, UK.

[2]School of Chemistry and Photon Science Institute, University of Manchester, Oxford Road, Manchester, M13 9PL, UK.

[3]Beijing National Laboratory of Condensed Matter Physics, Institute of Physics, Chinese Academy of Sciences, Beijing 100190, China

[4]Centre for Mesoscience and Nanotechnology, University of Manchester, Manchester M13 9PL, UK.

[5]Center for Nano-metrology, Korea Research Institute of Standards and Science, 267 Gajeong Ro, Yuseong-Gu, Daejeon, 305-340, Republic of Korea.

[6]Institute for Molecules and Materials, Radboud University of Nijmegen, Nijmegen 6525 AJ, The Netherlands.

[7]Institute of Quantum Materials Science, Ekaterinburg 620075, Russia

[8]National Institute for Materials Science, 1-1 Namiki, Tsukuba 305-0044, Japan

*When a crystal is subjected to a periodic potential, under certain circumstances (such as when the period of the potential is close to the crystal periodicity; the potential is strong enough, etc.) it might adjust itself to follow the periodicity of the potential, resulting in a, so called, commensurate state[1-3]. Such commensurate-incommensurate transitions are ubiquitous phenomena in many areas of condensed matter physics: from magnetism and dislocations in crystals, to vortices in superconductors, and atomic layers adsorbed on a crystalline surface[1]. Of particular interest might be the properties of topological defects between the two commensurate phases: solitons[2,4], domain walls[1], and dislocation walls[5-7]. Here we report a commensurate-incommensurate transition for graphene on top of hexagonal boron nitride (hBN)[8,9]. Depending on the rotational angle between the two hexagonal lattices, graphene can either stretch to adjust to a slightly different hBN periodicity (the commensurate state found for small rotational angles) or exhibit little adjustment (the incommensurate state). In the commensurate state, areas with matching lattice constants are separated by domain walls that accumulate the resulting strain. Such soliton-like objects present significant fundamental interest[1], and their presence might explain recent observations when the electronic, optical, Raman and other properties of graphene-hBN heterostructures have been notably altered[10].*

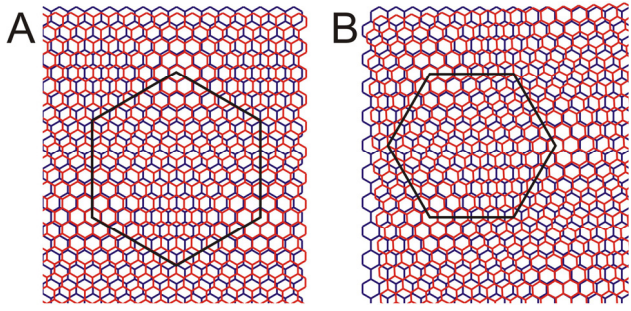

Fig. 1 Schematic representation of the moiré pattern of graphene (red) on hBN (blue) for φ=0° **(A)** and φ=3°≈0.052rad **(B)**. The mismatch between the lattices is exaggerated (~10%). Black hexagons mark the moiré plaquette.

The classical system which is used to simulate commensurate-incommensurate transitions is a one-dimensional chain of elastically linked atoms in a background periodic potential – the Frenkel-Kontorova model[2]. The two-dimensional (2D) version of the model[11,12] can be applied to real-life systems, such as the surface reconstruction at the interface between two crystals (or between a crystal and a surface monolayer) with close atomic lattice periods[4,6,7,13]. Commensurate-incommensurate transitions in 2D have been discussed[3,6,7,13] and observed[14] previously in systems such as adsorbed atoms on a surface of a crystal. Interestingly, the boundaries between the commensurate phases can be described in terms of topological defects. In the one-dimensional case such defects are usually described by solitons[1,2,4], whereas in 2D the language of misfit dislocations is commonly used[5]. The ultimate way to observe such reconstruction would be by monitoring the behaviour of two 2D atomic crystals when placed in close contact.

Recent advances in the production of heterostructures based on 2D atomic crystals[15], and, in particular, the preparation[8] and growth[16] of graphene on hBN, allow us to revisit this problem. hBN has been originally utilised as a substrate[8,17] and also an encapsulation layer[18], which allows minimisation of the detrimental influence of $SiO_2$ substrates, and, as a consequence, the achievement of spectacular electronic quality of the resulting graphene devices. Still, the van der Waals interaction between hBN and graphene , however weak, is not negligible (≈10 meV per carbon atom)[19]. The mismatch $\delta =a_{hBN}/a_G-1\approx 1.8\%$ between the lattice constants of hBN ($a_{hBN}$) and graphene ($a_G$) and the relative rotation angle φ between the graphene and hBN crystals lead to a hexagonal moiré pattern (Fig. 1), which has been observed by scanning probe microscopy[20,21]. The moiré potential acts on charge carriers in graphene resulting in a modification of its electronic spectrum[22-25]. So far it has been assumed that no structural changes occur in graphene after it is brought in contact with hBN.

In this paper we investigate (by atomic force microscopy (AFM), scanning tunnelling microscopy and Raman spectroscopy) the strain distribution in graphene on hBN for different misorientation angles between the crystalline structure of the two crystals (which results in the variation of the period $L$ of the moiré pattern). We observe a commensurate-incommensurate transition that occurs when φ is of the order of δ (that is, ≈1°). For φ<δ (large $L$ >10nm), graphene stretches locally to achieve an energetically favourable state for van der Waals interactions with hBN, which results in relatively large areas  of commensurate stacking and deformations concentrated in narrow strained regions (similar to soliton lattice formation in one dimension[1,2]). For φ>δ (small moiré periodicity), graphene and hBN lattices remain unsynchronized and there are no distinct regions with accumulated strain.

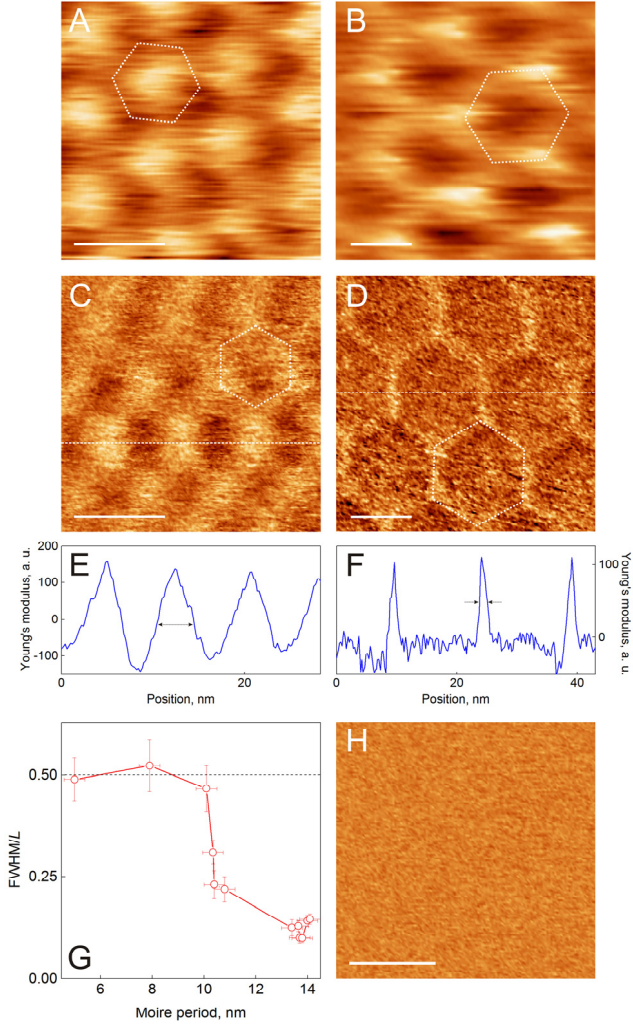

Fig. 2 (**A**) Local resistance measured by conductive AFM for one of our graphene-on-hBN samples with an 8 nm moiré pattern. Colour scale: white to black is 105 kΩ to 120 kΩ. (**B**) Same as in (A) for a sample with a 14 nm moiré periodicity – the crystallographic axes of graphene and hBN are practically aligned. Colour scale: from 135 kΩ to 170 kΩ. (**C** and **D**) Young modulus distribution, measured in the PeakForce mode, for structures with 8 and 14 nm moiré patterns, respectively. (**E** and **F**) Cross-sections of the Young modulus distribution taken along the dashed line in (C) and (D), respectively, and averaged over 10 scanning lines (approx. 2.5nm). (**G**) Ratio between FWHM of the peak in the Young's modulus distribution (as marked by dashed arrows in (E) and (F)) and the period of moiré structure *L*, as a function of the period of the moiré structure for several of our samples. (**H**) Young's modulus distribution across an unaligned sample (angle between graphene and hBN ~15°).

Our samples are fabricated by the dry transfer method described in detail in refs[9,18]. In brief, graphene is prepared by micromechanical cleavage[26] on top of a polymer film consisting of two sacrificial layers. By dissolving one of them, graphene supported by the second layer can be transferred on top of a relatively thick (>10nm) crystal of hBN. Then the second sacrificial layer is also removed and the assembly is annealed in a forming gas at temperature ~250°C to achieve an atomically clean interface[27]. We control φ with ~0.5° precision as described previously[22]. In some of the experiments described below, we have used structures with another hBN crystal (1-5nm thick) added on top to encapsulate graphene. The quality of the graphene-hBN interface and the presence of the moiré pattern have been confirmed by transport measurements[22] and by conductive AFM measurements[22].

We studied our samples by AFM (in various modes) and Raman spectroscopy. No moiré pattern has been observed in AFM topography signal - neither in contact nor in tapping modes (precision ~50pm). At the same time, the moiré pattern can be seen in the friction signal, which may indicate areas with different adhesion[28]. To elaborate on this finding we have measured force curves at different positions on the surface. To do that we have utilised PeakForce Tapping AFM[29], which allows us to extract local elastic constants including the Young modulus and adhesion[30,31].

First we investigate a structure with φ~1.5° and *L* = ~8nm (see Figs 2A, C, E). The moiré pattern is clearly seen in the conductive AFM (Fig. 2A) and Young modulus signals (Fig. 2C). There is no signal in adhesion channel.

The pattern of Young modulus distribution for the *L* = 8 nm sample has a hexagonal symmetry with smooth changes across each moiré unit cell. Fig. 2E shows that the cross section is close to the sinusoidal shape. Overall, it follows the prediction of Sachs *et al*[19], where both the Young modulus and the adhesion follow a characteristic hexagonal structure. The reason we do not see any pattern in the adhesion signal is probably because the adhesion between graphene and hBN is much larger than the adhesion between the AFM tip and graphene.

The situation is very different for the sample with φ~0° and *L* = 14 nm. The pattern of the Young modulus signal is still hexagonal, but it changes abruptly in space (Fig. 2D), with large areas of low, practically constant Young modulus separated by narrow regions of high modulus (domain walls, seen as bright lines on Fig. 2D). The cross section reveals that the latter is only ~2nm wide. Similar behaviour is observed by conductive AFM: smooth variations for 8 nm moiré samples and sharp features separating domains of practically constant conductivity in the 14 nm moiré patterns. Note, that the hexagonal symmetry is broken, with every second vertices being much brighter. It might indicate a specific stacking between graphene and hBN or a concentration of strain, which alters the electronic structure through the effect of pseudo-magnetic field[32,33].

We summarise this finding in Fig. 2G, where we plot the ratio between the FWHM of the peak in the Young modulus to the period of the moiré pattern for the same sample. For samples with *L*<10nm this ratio saturates at 0.5 – as it should be if there is no reconstruction of the graphene lattice, Fig. 2E. For our best aligned samples (*L*~14nm), this ratio is close to 0.1. The relative width of the peak in the Young modulus gradually grows as the alignment becomes less perfect and, at *L*~10nm, the pattern suddenly becomes practically sinusoidal (the ratio is ~0.5).

Such a change in behaviour can be explained by the commensurate-incommensurate transition as a function of φ. The basic physics behind such transitions is as follows[1,3,4]: if the relative rotation angle between the two crystals is small (large period of the moiré pattern), it becomes energetically favourable to adjust the two lattices to become commensurate, losing in elastic but gaining in the van der Waals energy. The latter decreases if preferred atomic positions are achieved over the whole area. When φ increases past some critical value (so that the period of moiré pattern become small), the gains in the van der Waals energy can no longer compensate for the elastic energy and the two crystals act independently, forming an incommensurate state.

In principle it is possible to imagine a situation when the commensurate state would extend across the whole interface between the two crystals and the crystals would be uniformly stretched or compressed (it would happen, for instance, if the gain in van der Waals energy is sufficiently large and in the absence of the 3D elastic fields in the substrate). In our case, however, when graphene is mechanically deposited on hBN, such uniform stretching of graphene would require its macroscopic motions. In practice, graphene is always pinned by imperfections, so its size is fixed. This leads to the formations of domains (where graphene and hBN are commensurate) and domain walls (where graphene is compressed and the stacking order changes rapidly in space). Note, that in such a scenario the overall period of the moiré pattern doesn't change, but the strains change sign in space.

In the Frenkel-Kontorova model the adjacent regions of commensurate phase (when the atoms falls into the minima of the background periodic potential – the, so called, Peierls potential[2]) are separated by solitons. The width of the soliton is given by $\lambda \propto (Y/\gamma)^{1/2}$ where *Y* is the Young modulus and γ is the depth of the background periodic potential. The commensurate-incommensurate transition (although it is not necessarily a phase transition in one-dimensional model) occurs when the size of the soliton becomes comparable with the size of the commensurate region, such as for very stiff crystals (large *Y*) or for very shallow background potential (small γ).

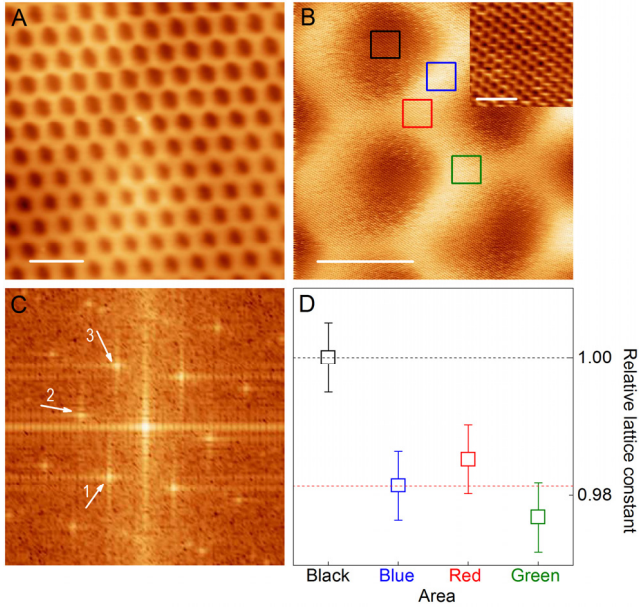

Fig. 3 (**A**) STM image of one of our aligned samples. Moiré pattern is clearly visible. Scale bar 30nm. Sample bias -0.1V, tunnelling current 300pA. (**B**) Same as in (A), just under higher magnification. Both the moiré pattern and the atomic structure are resolved. Scale bar 10nm. Sample bias -0.1V, tunnelling current 800pA. Coloured squares (3nm in size) indicate the fragments used for Fourier transformation to determine the interatomic distance. Inset: a blow up of the area marked by the black square. Atomic structure is clearly visible. Scale bar – 1nm. (**C**) Example of the Fourier transformation of atomically resolved structure. In this case, as the starting image we used 3nm × 3nm square image at the vertex of the hexagonal pattern (red square in B). Scale bar 5nm$^{-1}$. (**D**) Relative lattice constants (with respect to those measured for the area marked by black square in (B)) for different areas within the moiré pattern (colours corresponds to those in (B)). Obtained from the positions of the first order peaks in (C) and averaged over the three directions.

The 2D case of graphene on hBN can be qualitatively traced back to the Frenkel-Kontorova model with the role of solitons (topological defects) being played by a system of screw and edge dislocations (the former are associated with rotations whereas the latter are introduced by a lattice misfit)[5,34] between graphene and the hBN lattice. Note that even though the language of dislocations is used here, graphene is still defect-free and the dislocations reflect only the mismatch in the lattice constants of graphene and hBN in a certain direction. As the rotation angle increases, two processes occur: (i) the distance between the dislocations becomes smaller (shortening of the period of the moiré pattern) and, (ii) the width of the dislocation cores increases due to a flattening of the effective Peierls potential. The latter is illustrated in Fig. 1: for φ=0 the moiré plaquette is aligned with crystallographic directions of graphene and hBN and thus the Peierls potential is atomically sharp. At the same time, for any finite φ the moiré plaquette is misaligned with respect to the crystallographic directions, thus the Peierls potential has a complicated shape, with a period much larger than the lattice period. As a result, when the rotation angle reaches the value of the order of the lattice misfit (0.018) the dislocation cores become as large as the moiré period itself. This is nothing but the commensurate-incommensurate transition: for smaller rotation angles, all lattice misfit is concentrated in relatively narrow, well-defined dislocation walls, whereas for larger misorientation, the angular and lattice misfit is more or less uniformly distributed through the whole system. Detailed quantitative analysis demonstrate that the switchover between commensurate and incommensurate phases indeed happens through a phase transition[34].

To test this theory we directly measured the interatomic distances at different parts of the moiré pattern with scanning tunnelling microscopy, Fig. 3. STM image shows reconstructed moiré pattern (narrow domain walls like objects) on all 3 tested samples, which is in agreement with previous STM data[20,21]. We analysed the interatomic distances at different areas of the moiré pattern (Fig. 3B), namely: at the body of the hexagon (dark areas in Fig. 3A, B, marked by black square in Fig. 3B), at the vertexes of the hexagons (where the three domain walls merge, marked by red and green squares in Fig. 3B) and in the middle of the domain walls (blue square in Fig. 3B). The interatomic distance was analysed by taking a 2D Fourier transformation and observing the positions of the first order peaks. We took care to compare the positions of the peaks which correspond to the same crystallographic directions (Fig. 3C). This way we avoided the artefacts associated with the thermal drift.

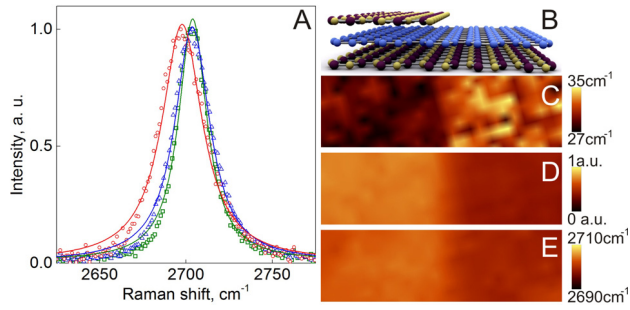

Fig. 4 **(A)** Raman 2D peak for various graphene samples on hBN samples. The data is normalised by the amplitude of the 2D peak. Symbols are the experimental data points, with lines of the corresponding Lorentzian fit in the respective colour. Red circles and line – for graphene on hBN with $L \approx 14$ nm, non-encapsulated. Blue triangles and line – for the same $L$ but encapsulated. Green squares and line - $L \approx 8$ nm, non-encapsulated. **(B)** Schematic representation of one of our devices with graphene partially encapsulated. **(C)** Spatial map of the FWHM of 2D peak for a graphene on hBN sample with the period of moiré pattern 14nm. Left half of the sample is encapsulated with a few layers of hBN. Crystallographic directions of the top hBN are rotated by approximately 15º with respect to those of graphene. **(D)** Spatial map of the amplitude of the 2D peak for the same sample as in (C). Normalised to the amplitude of the G peak. **(E)** Spatial map of the position of the 2D peak for the same sample as in (C). Laser excitation is 488nm for all Raman data.

Throughout all our samples we found that the interatomic distances within the body of the domains are consistently larger than that within the domain walls, Fig. 3D. The difference is 2.0%±0.6%, see[34] for further details. The sign and the value of the effect are consistent with the above theory. The fact that the lattice extension with the domain area (marked by black square in Fig. 3B) with respect to the other parts of the moiré pattern comes larger than δ suggests that the lattice within the domain walls (blue, red and green areas in Fig. 3B) is most probably compressed. Unfortunately, at this stage we can't say anything about the specific direction of the strain within the domain walls, as we are working at the limit of the resolution of our STM.

It is clear that the strain distribution in graphene on hBN is quite different for commensurate and incommensurate states. Recently, it has been demonstrated that the FWHM of the Raman 2D peak for graphene with $L = 14$ nm is about 50% larger than that for samples with the 8 nm moiré pattern[10] (the result is reproduced in Fig. 4A). Such behaviour could be explained by elastic deformations in the aligned graphene[35].

One can expect that the commensurate state of graphene on hBN can be suppressed by placing an additional hBN crystal on top. If the top hBN is rotated by a large angle ($\varphi>5º$), graphene would experience an additional van der Waals potential with a short $L$[20]. This should reduce the influence of the bottom van der Waals potential and the combined effect may lead to the disappearance of the commensurate state.

We have checked this behaviour by producing encapsulated graphene samples where graphene is aligned relative the bottom hBN, and the top hBN rotated by approximately 15 degrees with respect to the crystallographic directions of graphene (top hBN covers graphene only partially). It is impossible to observe the moiré pattern by AFM on the covered part of graphene, so only the comparative study of the Raman signal on the covered and uncovered parts of graphene has been performed. The uncovered graphene has been found to be in commensurate state as confirmed by both AFM and Raman. As for the covered graphene, the Raman peak is found not to be broadened in comparison with graphene on hBN with 8 nm moiré pattern. This means that encapsulated graphene remains in an incommensurate state even if thoroughly aligned.

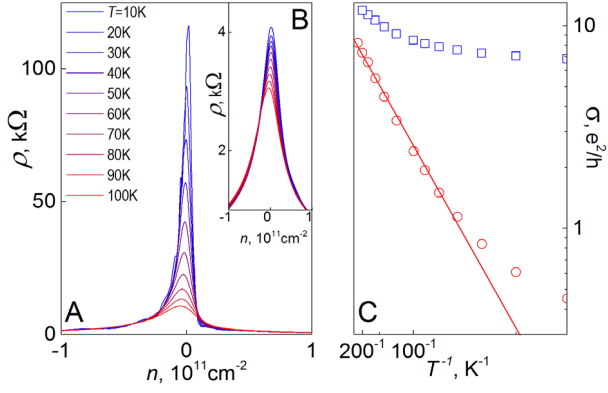

Fig. 5 Longitudinal resistivity as a function of carrier concentration for **(A)** non-encapsulated and **(B)** encapsulated graphene on hBN with the 14 nm moiré pattern. **(C)** Temperature dependence of the conductivity minimum for the samples in (A) – red circles and in (B) – blue squares. The red line is a guide for an eye, yielding $\Delta/2 \approx 180K$.

The observed strain in graphene aligned on hBN and the quenching of the commensurate state in encapsulated aligned samples can explain recent observations of a gap opening in some graphene-on-hBN devices[24] and its absence in others[8,17,18]. The difference between the two sets of devices was the hBN crystals put on top of the latter devices. To confirm this hypothesis, we have prepared similar sets of aligned graphene devices and studied their transport properties. The devices were standard Hall bars, and $L$ was approximately the same (14 nm) as found from the gate voltage at which the secondary Dirac points occur[22].

In the non-encapsulated devices the commensurate transition has been observed by AFM and Raman spectroscopy. These devices exhibit an insulating behaviour at the main neutrality point (Fig 5A). The associated gap $\Delta$ is estimated as $\approx 360K$ by fitting the high temperature data with the Arrhenius law, Fig. 5C. At temperatures below 60 K the insulating behaviour shows a slower dependence, which may indicate the onset of hopping conductivity. The size of the gap is similar to that reported previously for aligned but not encapsulated graphene[24]. In contrast, our encapsulated samples, which are identified by Raman spectroscopy as being in the incommensurate state, exhibit weak temperature dependence with resistivity of the order of several k$\Omega$ at low temperatures, Fig. 5B. This leaves a possibility of only a small gap at the main neutrality point, much smaller than that observed in non-encapsulated devices. Furthermore, we have studied tens of graphene-on-hBN devices (encapsulated and open) and never observed a gap in those with $L$ <10 nm ($\varphi \sim 1^{\circ}$). .

Therefore, the gap at the main Dirac point can be associated with the commensurate state. The sublattice symmetry in graphene is locally broken due to the proximity to hBN. However, the resulting local gaps vary spatially, and the global transport gap may be small due to averaging[19,36]. In the commensurate state, large areas of graphene would have the same crystal structure as hBN and, therefore, a constant magnitude of the gap. This would strongly enhance the global transport gap and can be responsible for the observed large $\Delta$. An alternative explanation would be that the transport is limited by percolation through the system of the "domain walls". The insulating behaviour in such regions (Fig. 2D), can be due to a strong inhomogeneous strain that leads to energy gaps often interpreted in terms of large pseudo-magnetic fields[32,33,37].

Finally, we would like to discuss the possible microscopic strain distributions for our samples in commensurate state. Two structures for the boundary between adjacent commensurate domains are possible: those which accommodate tensile, and shear strain. The shear strain is, however, more energetically favourable as the shear modulus is half the Young modulus[38]. This observation is also supported by our Raman measurements: a tensile-type of strain distribution would require a very large strain accumulated within the narrow boundary (of the order of 10%) which would be observed as much larger broadening of the Raman 2D peak. Similar conclusions can be drawn from the STM results: a tensile-type of strain distribution would result in much larger difference in the lattice constants between the domains and the domain walls. The possibility of generating a periodic distribution of shear strain allows for local strain concentration and calls for further study.

In conclusion, graphene placed on hBN experiences a commensurate-incommensurate transition for the period of moiré pattern larger than ~10nm. The formation of large commensurate regions is associated with creation of transition regions (domain walls, where strain is accumulated) and results in broadening of the Raman 2D peak and appearing of an insulating state at the Dirac point. Such topological defects, with nontrivial strain distributions, might be used to modify the electronic states via pseudo-magnetic field.

This work was supported by the European Research Council, Graphene Flagship, Engineering and Physical Sciences Research Council (UK), the Royal Society, U.S. Office of Naval Research, U.S. Air Force Office of Scientific Research, U.S. Army Research Office, the MOST of China (No. 2013CBA01600) and the Körber Foundation. We are grateful to L. Levitov for useful discussions.

# Commensurate/incommensurate transition for graphene on hexagonal boron nitride

C. R. Woods[1], L. Britnell[1], A. Eckmann[2], R. S. Ma[3], J. C. Lu[3], H. M. Guo[3], X. Lin[3], G. L. Yu[1], Y. Cao,[4] R. V. Gorbachev[4], A.V. Kretinin[1], J. Park[1,5], L. A. Ponomarenko[1], M. I. Katsnelson[6], Yu. N. Gornostyrev[7], K. Watanabe[8], T. Taniguchi[8], C. Casiraghi[2], H. -J. Gao[3], A. K. Geim[4], K. S. Novoselov[1]

[1]School of Physics and Astronomy, University of Manchester, Manchester M13 9PL, UK.

[2]School of Chemistry and Photon Science Institute, University of Manchester, Oxford Road, Manchester, M13 9PL, UK.

[3]Beijing National Laboratory of Condensed Matter Physics, Institute of Physics, Chinese Academy of Sciences, Beijing 100190, China

[4]Centre for Mesoscience and Nanotechnology, University of Manchester, Manchester M13 9PL, UK.

[5]Center for Nano-metrology, Korea Research Institute of Standards and Science, 267 Gajeong Ro, Yuseong-Gu, Daejeon, 305-340, Republic of Korea.

[6]Institute for Molecules and Materials, Radboud University of Nijmegen, Nijmegen 6525 AJ, The Netherlands.

[7]Institute of Quantum Materials Science, Ekaterinburg 620075, Russia

[8]National Institute for Materials Science, 1-1 Namiki, Tsukuba 305-0044, Japan

## Commensurate-incommensurate transition

Conditions of conjugation of two half spaces in a crystal have been discussed long ago in the context of grain boundaries [1,2]. The situation when a half of space is rotated (at angle $\phi$) with respect to another half of space is called a twist grain boundary. There are two canonical ways to describe this situation. The first one is the simplest picture of a coincidence site lattice (CSL) [2] where one just rotates and puts one lattice onto the other without atomic relaxation; it corresponds to the moiré description used in previous works on graphene on hexagonal boron nitride (hBN) [3-6]. The second approach is based on the theory of dislocations [1]. Namely, one introduces two systems of screw dislocations with their axes (and, therefore, Burgers vectors $\vec{b}$ which, for screw dislocations, are parallel to the axes) normal to the rotation axis. After this, the interface should be reconstructed to minimize the total energy. This reconstruction is well known for the one-dimensional case of two interacting chains with the lattice misfit where it results in a formation of soliton lattice [7,8]. This language is frequently used for a monolayer of adsorbed atoms rotated with respect to the host crystal [9-12].

The moiré pattern typical for the graphene-hBN system [3-6] arises naturally for a completely different situation, namely, in computer simulations for metals, see, e.g., Fig.5 of Ref. [10]. The description of the moiré pattern by the wave vector $\vec{k} = \vec{g} - \vec{g}'$ where $\vec{g}, \vec{g}'$ are the shortest reciprocal lattice vectors of graphene and hBN, respectively [3] agrees naturally with the dislocation picture. The main difference between these two languages is that CLS assumes a matching of the layers only for special sites with the period $2\pi/k$, whereas in the dislocation pictures it takes place almost everywhere, except the dislocation cores. Thus, the description of rotated layers in terms of dislocations corresponds to *relaxed* moiré [9].

Keeping in mind applications to graphene-hBN, we consider only the case when the energy of interlayer (van der Waals) interaction $V$ is much weaker than the cohesive energy within the layer and the misfit of lattice constant $b-a$ is small in comparison with $a$ and $b$. Then, the interface structure results from an interplay of the van der Waals and elastic energies which are characterized by the parameters $V$ and $E_{el} \approx Y(b-a)^2/2$, respectively ($Y$ is the in-layer Young modulus). When $E_{el} >> V$ the CSL picture (a naïve, that is, unreconstructed, moiré picture) with two incommensurate periods is energetically favorable but in the opposite case $E_{el} << V$ the lattice constants want to be equal to optimize the van der Waals energy, all lattice misfit being focused into "soliton lattice" with large commensurate regions between narrow solitons [8]. The "commensurate-incommensurate transition" takes place at $E_{el} \approx V$. The dislocation picture described above is a two-dimensional analog of the soliton lattice regime, and the commensurate-incommensurate transition at the change of misfit has been observed in many simulations for an adsorbed monolayer at metal and semiconductor surfaces [9-12]. For graphene-hBN system a typical energy of van der Waals interactions estimated as a difference between the total energies for A-B and A-A stacking corresponds to 20 meV and is slightly larger than the corresponding energy of elastic distortions involved in the formation of a moiré pattern [13]. In terms of dislocations, the "commensurate" ("soliton lattice") phase corresponds to the situation of narrow dislocation cores, the core size is much smaller than the interdislocation distance. The commensurate-incommensurate transition corresponds to the case of overlapped cores when the description in terms of dislocations becomes meaningless. This can happen for both the decrease of the interdislocation distance and the increase for the core width; as we will see below both factors are relevant.

For triangular Bravais lattices the twist of top layers can be described by a superposition of two arrays of screw dislocations with the Burgers vectors $\vec{b}_1 = a(1,0)$, $\vec{b}_2 = a/2(1,-\sqrt{3})$ as shown in Supplementary fig. 1a, where $a$ is the lattice constant. The distance between dislocations in each array is $\lambda = b/\left(2\sin\frac{\phi}{2}\right)$, where $\phi$ is the rotation angle. Note that for the case of tilt grain boundaries one has to introduce a family of edge dislocations, with the same expression for the interdislocation distance (in that case $\phi$ is the tilting angle) [1,2]. Reaction between crossed dislocations $\vec{b}_1 + \vec{b}_2 \to \vec{b}_3$ is energetically favorable [1] and results in a formation of the hexagonal network (Supplementary fig. 1b); each segment of the network represents the screw dislocation, and the conservation equation for Burgers vector takes place in each triple joint. This dislocation networks (we will call them twist dislocations) determine a (plastic) rotation of the upper layer with respect to the lower one, the inner region of each hexagon being rotated in the opposite direction providing a complete matching of the layers. Alternative dislocation description was proposed in Ref.[9]; in that model, the conservation of the Burgers vector for each node is provided by involvement of threading (vortex) dislocations resulting in formation of so called "bright stars" and "herringbone" patterns.

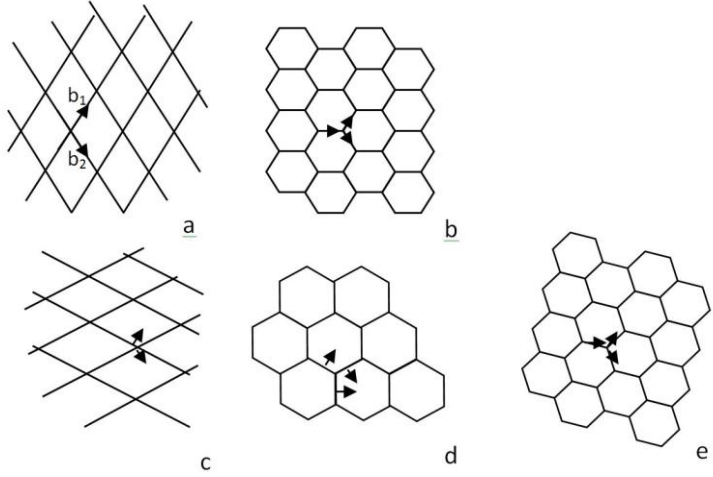

*Fig. S1 - Dislocations arrays providing a rotation (a,b) and compensation of lattice misfit (c,d) for the graphene layer on hexagonal boron nitride. Lines correspond to dislocation axes and arrows to the Burgers vectors. Reactions between screw (a) or edge (c) dislocations $\vec{b}_1 + \vec{b}_2 \to \vec{b}_3$ lead to a formation of hexagonal networks (b) and (d). The hexagonal network at the figure (e) is obtained by a rotation of the network (b) by the angle $\alpha$ (S2) with the same directions of the Burgers vectors and provides both the twist of graphene layer and compensation of the lattice misfit everywhere except the dislocation walls forming the network.*

To compensate the lattice misfit between graphene and hBN, two families of edge dislocations with the axes perpendicular to the Burgers vectors $\vec{b}_1$ and $\vec{b}_2$ are required (Fig. S1c). Reactions between these dislocations also lead to the formation of a hexagonal network (Fig. S1d), which we believe is clearly seen at Fig. 1D of the main text. The complete set of the misfit dislocations can be represented as a superposition of the families of screw (Fig. S1b) and edge (Fig. S1c) dislocations. Numerical simulation [10,12] shows that this superposition results in a hexagonal network of dislocations obtained by a rotation of the network of twist dislocations by the angle of rotation of the dislocation network $\alpha$ (which is defined explicitly below), keeping unchanged the directions of the Burgers vector with respect to the graphene lattice. The resulting network (Fig. S1e) consists of the dislocations with both screw and edge components providing both rotation and compensation of the lattice misfit.

Let us consider the geometry of the dislocation network displayed in supplementary fig. 1d which is formed by two crossing families of dislocations with axes $\xi_1$, $\xi_2$ rotated with respect to the Burgers vectors by the angle $\alpha$. If $\alpha$ is different from 0 (screw dislocations) and $90^0$ (edge dislocations) both screw and edge components are present. The tensor of plastic deformation for each family reads [1]

$$\beta_k = \begin{pmatrix} 0 & \beta_{\xi_k n} \\ 0 & \beta_{nn} \end{pmatrix}, \qquad (S1)$$

where $\vec{n}$ is the vector normal to the dislocation axis, $\beta_{\xi n} = b_\xi / \lambda = b\cos(\alpha)/\lambda$ is the average shear deformation and $\beta_{nn} = b_n / \lambda = b\sin(\alpha)/\lambda$ is the uniaxial expansion/compression. The antisymmetric part of the tensor $\beta$ determines a rotation of the upper layer with respect to the lower one by a vector $\vec{\omega}$ whereas its symmetric part provides a compensation of the lattice misfit $\delta$

$$|\vec{\omega}| = \frac{b\cos(\alpha)}{\lambda}, \qquad \delta = \frac{b\sin(\alpha)}{\lambda} \tag{S2}$$

At the absence of rotations ($\phi = 0$, $\alpha = 90^0$) the inter-dislocation distance is $\lambda = b/\delta$. Oppositely, for a pure rotation of identical layers $2\sin(\phi/2) = b/d$. At $\delta \neq 0$ additional rotation is required at the angle α determined by the relation

$$\tan(\alpha) = \frac{\delta}{2\sin(\phi/2)}, \tag{S3}$$

and for the period of the network one has (cf. Ref. [3], Supp Info)

$$\lambda = \frac{b}{\sqrt{\delta(\delta + 2\sin(\phi/2))}} \tag{S4}.$$

The geometry of the dislocation network is similar the moiré pattern, but the angle α is related with the angle θ of rotation of the moiré pattern used in Ref.[3], Supp Info, Eqs.(3)-(7), by the relation $\theta = 90^0 - \alpha - \phi$. For graphene-hBN $\delta = (b-a)/a \approx 0.018$ and at $\phi \approx \delta$ the angle θ becomes large. For example, at $\phi = 2.5^0$ the angle θ = $60.2^0$ and λ=30a. The essential difference is that in our description all misfit is concentrated in the dislocation cores with a complete lattice conjugation within the hexagonal cells.

To discuss possible reconstructions of the dislocations at the increase of the angle $\phi$ we need to build an effective one-dimensional potential relief V(u), u is the relative displacement of atomic rows parallel to the dislocation axis. For simplicity, let us consider the case of commensurate periods of superlattice and basic lattice along the dislocation line, with rational $\tan\theta/2$, which leads to rational $\cos\theta$ and $\sin\theta$ and, thus, to commensurability of the dislocation network with crystal lattices. Let integer M be the number of atoms in the dislocation line per elementary cell of the superlattice. The coordinates of these atoms are $\vec{r}_m = ma\vec{e}$ where $m = 0,...,M-1$ and $\vec{e} = (\cos\theta, \sin\theta)$ is the unit vector along the dislocation line. For atomic displacement along the line $u\vec{e}$, the potential relief is equal to:

$$V(u) = \frac{1}{M}\sum_{\vec{R}}\sum_{m=0}^{M-1}\varphi(\vec{r}_m - \vec{R} - u\vec{e}) = \frac{1}{M}\sum_{\vec{g}}\varphi_{\vec{g}}\sum_{m=0}^{M-1}e^{i\vec{g}\vec{e}am}e^{i\vec{g}\vec{e}u} = \frac{1}{M}\sum_{\vec{g}}\varphi_{\vec{g}}\frac{1-e^{i\vec{g}\vec{e}aM}}{1-e^{i\vec{g}\vec{e}a}}e^{i\vec{g}\vec{e}u} \tag{S5}$$

where $\varphi(\vec{r} - \vec{R})$ is an interatomic potential between the upper and lower layers, $\varphi_{\vec{g}}$ its Fourier component, $\{\vec{R}\},\{\vec{g}\}$ are direct and reciprocal crystal lattices for the lower layer. In the two-dimensional Frenkel-Kontorova model [14] we can take into account only minimal $\vec{g}$-vectors, which are, for triangular Bravais lattice,

$$\vec{g}_1 = \frac{2\pi}{a}(1,0), \quad \vec{g}_2 = \frac{\pi}{a}(-1,\sqrt{3}) \tag{S6}$$

One can see that for small $\alpha$ the potential (S5) is close to a pure cosine whereas for large enough θ it contains several harmonics with comparable coefficients. The shape of oscillations of the potential relief is essentially dependent on θ and M. To illustrate this, we draw just a contribution proportional to $\varphi_{g_1}$. Supplementary fig. 2 shows an envelope of oscillations as a function of $\theta$ for different M. One can see that the amplitude of the oscillations drops sharply at $\theta \approx 12-15^0$ decreasing in more than an order of magnitude for $\theta \approx 25^0$ and values of M typical for the case of graphene-hBN (see Fig. 1 of the main text).

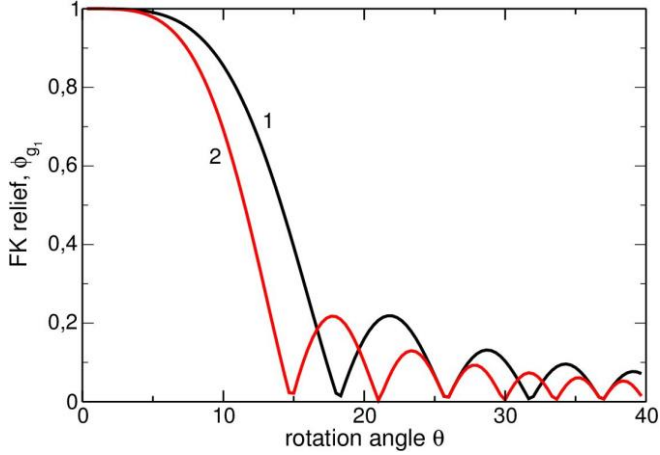

Fig S2 - Dependence of the envelope of potential Frenkel-Kontorova (FK) relief (in the units of $\varphi_{g_1}$) on the pattern rotation angle $\theta$ at M = 20 (curve 1) and 30 (curve 2).

Within the Frenkel-Kontorova model, the atomic displacements are given by the expression

$$u(x) = 4\arctan\left[\exp\left(\frac{x-x_0}{\sqrt{K}}\right)\right] \quad (S7)$$

where $K \propto Y/\gamma_{un}$ (Y is the Young modulus, $\gamma_{un} = V_0/h$ is the unstable stacking fault energy per unit area, $V_0$ is the amplitude of the potential relief and h is the period in the direction perpendicular to the dislocation axis) [15] which determines the width of the edge dislocation core as

$$\lambda = \frac{a}{2\pi}\sqrt{\frac{Y}{\gamma_{un}}} \quad (S8)$$

(for screw dislocation, Y is replaced by the shear modulus μ). According to the computations, Y = 22 eV/Å$^2$ [16], and, at $\phi=0$, $\gamma_{un}$ = 0.004eV/Å$^2$ [13], thus, $\lambda \approx 10a \approx 2.4nm$. After rotation of the graphene layer at $\phi=2°$ the dislocation network twists at $\theta \approx 25°$ the quantities $V_0$ and $\gamma_{un}$ are decreased by an order of magnitude (see Fig. S2) and $\lambda \approx 30a \approx 7.2nm$. Note that in this case $\xi \approx \lambda$ and the whole picture of individual dislocations is no more valid. We believe, however, that this approach is sufficient to predict the breakdown of commensurability and estimate its conditions.

Thus, a smooth increase of misalignment angle $\phi > \delta$, via dramatic increase of the angle $\theta$ can result in both decrease of the superlattice period $\lambda$ and increase of the dislocation-core width, leading to core overlap and commensurate-to-incommensurate transition.

## Other examples of samples in commensurate state

Overall we studied over a dozen samples. Below we demonstrate several examples of other device in commensurate state.

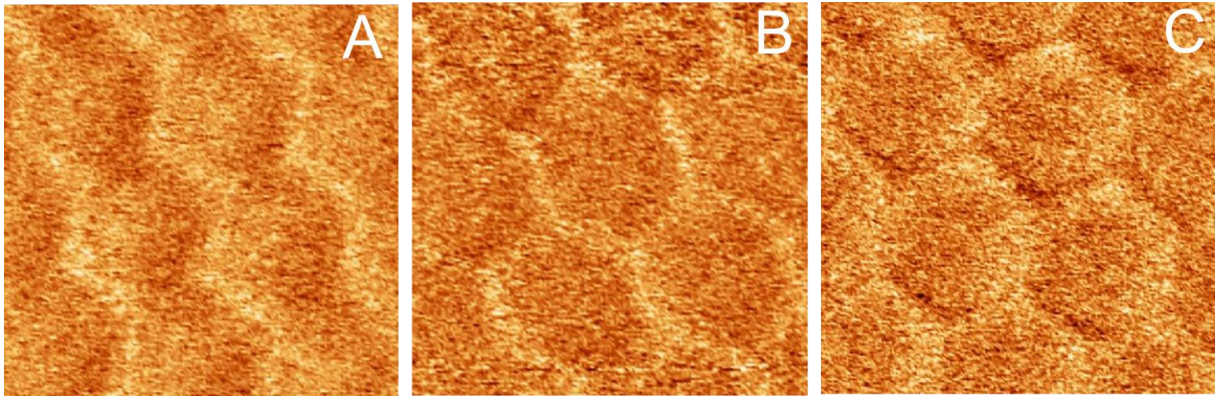

Fig. S3 – **(A)**, **(B)** and **(C)** are moiré patterns in Young's modulus for three further fully aligned samples. Each image is 40 x 40 nm.

## Details of Scanning tunnelling microscopy (STM) imaging

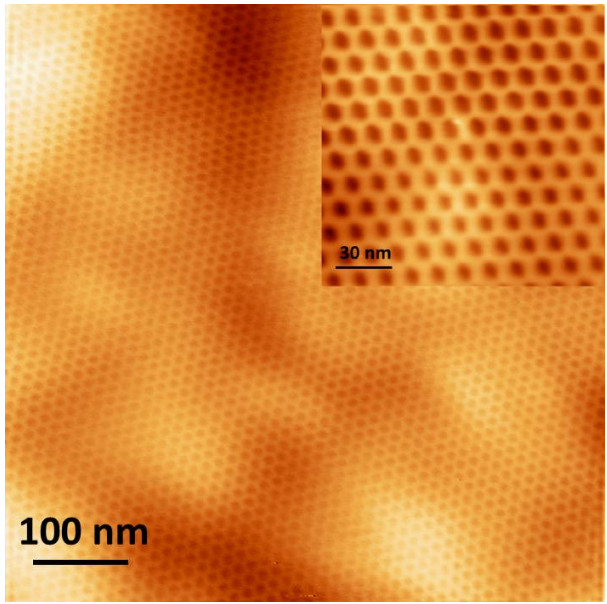

*Fig. S4 – moiré superlattice with a 14 nm period. (Inset) – An enlarged region of the superlattice.*

Images were taken of the hBN-graphene superlattice using STM imaging. Fig. S4 shows an image of the moiré superlattice.

Atomic resolution images of the moiré pattern were taken, to establish the variation of graphene's lattice constant across the superlattice period. Fig. S5, demonstrates an atomic resolution image of the pattern.

Fourier transforms of the graphene lattice at various points are taken to compare the graphene lattice constants variation across the superlattice period. Fig. S5 indicates the four regions which are used to extract the Fourier transform. Fig. S6 shows the Fourier transformed images for each of the regions labelled in Fig. S5. Three different directions were used to extract the relative values of graphene's lattice constant for each region. Only relative values are used because thermal drift distorts the observed graphene lattice. As shown in Fig. S6E,F,G the graphene lattice constant at region A is larger than that of regions B, C, and D.

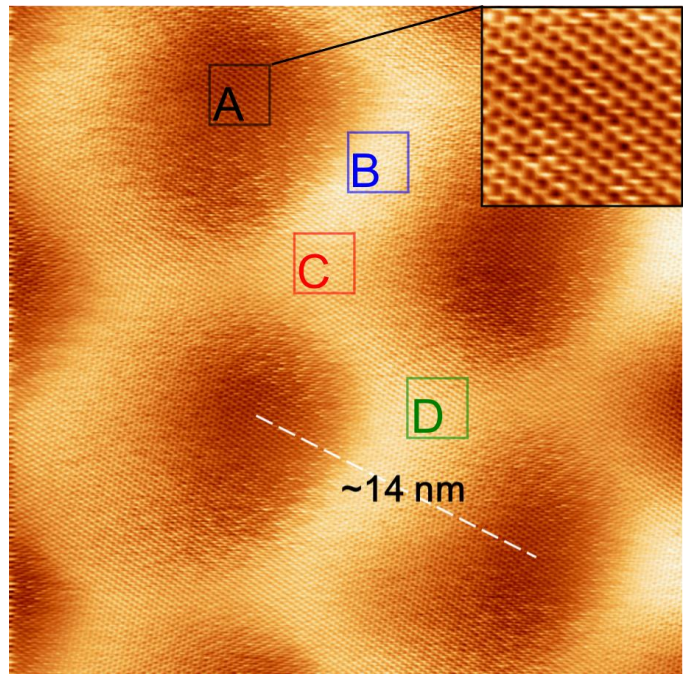

*Fig. S5 – A 30x30 nm atomic resolution image of the moiré pattern. **(Inset)** – expanded image of region 1 within the moiré pattern (3x3 nm). Fourier transforms of the regions labelled A, B, C, and D yield the graphene lattice constant at each of these points.*

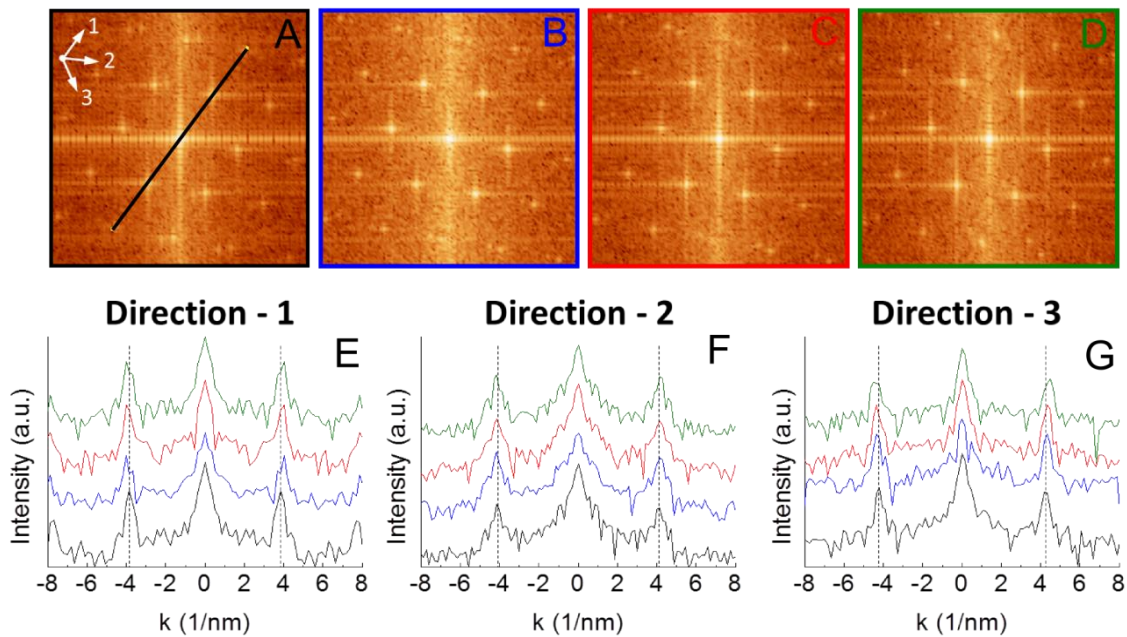

*Fig. S6 – **(A-D)** are Fourier transforms of regions **A, B, C,** and **D**, (as labelled previously in supplementary figure 5). **(Inset A)** indicates the directions in which the graphene lattice constant is measured in Fourier space. **(E), (F),** and **(G)** show the signal across the Fourier space in each direction. Dashed lines indicate the peaks corresponding to region A.*